\documentclass[epj,final]{svjour}
\usepackage{epsfig}
\usepackage{graphicx}
\begin{document}
\title{Nuclear and partonic dynamics in the EMC effect}
\author{Carlos A. Garc\'{\i}a
Canal\inst{1} \and Tatiana Tarutina\inst{1} \and Vicente Vento\inst{2}
}                     
%
%
\institute{Laboratorio de F\'{\i}sica Te\'{o}rica, Departamento de
F\'{\i}sica, IFLP, CONICET,\\ Facultad de Ciencias Exactas, Universidad
Nacional de La Plata
\\C.C. 67, 1900 La Plata, Argentina. \and Departamento de Fisica Teorica and Instituto de F\'{\i}sica Corpuscular,
Universidad de Valencia-CSIC, E-46100 Burjassot (Valencia), Spain.}
\date{Received: date / Revised version: date}
%
%
\abstract{
It has been recently confirmed that the magnitude of the EMC effect measured in electron deep inelastic scattering  is linearly related to the Short Range Correlation scaling factor obtained from electron inclusive scattering.  By using a $x$-rescaling approach we are able to understand the interplay between the quark-gluon and hadronic degrees of freedom in the discussion of the EMC effect.
} 
\maketitle

\section{Introduction}
\label{intro}

Deep Inelastic Scattering (DIS) provides a tool for probing the quark momentum distribution in
nucleons and in nuclei. Since the first indications that DIS structure functions measured in charged-lepton scattering
off nuclei differ significantly from those measured in isolated nucleons \cite{Aubert:1983xm,Arnold:1983mw} there has been
a continuous interest in fully understanding the microscopic mechanism responsible for the so called EMC effect and how it affects the 
momentum distribution of quarks in nuclei.

The experiment  E03-103  at  Jefferson Lab  has provided precise measurements of the EMC effect at large $x$,  in light nuclei, 
 $^3$He, $^4$He, $^9$Be and $^{12}$C \cite{Seely:2009gt}. In Refs. \cite{Higinbotham:2010tb,Weinstein:2010rt,Piasetzky:2011zz,Arrington:2012ax,Hen:2012fm}
 it was  shown that the EMC effect is linearly related to the short-range correlation (SRC) scale factor $a_2(A/d)$. This factor follows
 from the experimental observation that the ratio of inclusive electron scattering cross section from target nucleus $A$ and from Deuteron 
 scales for the Bjorken scaling variable   $x$, where $x=Q^2/2m_p q_0$, in the range $1.5 \le x \le 2$ at moderate $Q^2$  and therefore it has been suggested  
 \cite{Frankfurt:1993sp},  that
 
 \begin{equation}
 \sigma_A (1.5 \le x\le 2,Q^2) =\frac{A}{2} a_2(A/d) \; \sigma_2 (A, x,Q^2), 
 \end{equation}
 where $\sigma_2 (A,x,Q^2)$ is the effective cross section for scattering off a correlated $2N$ cluster in nucleus A.
 The SRC represent the high momentum components of the nuclear wave function, which can be described in terms of 
 nucleonic degrees of freedom. They are responsible  (in medium and heavy nuclei) for around $60$ \%  of the kinetic  energy 
 of nucleons in the nuclei \cite{Frankfurt:2009vv}. Assuming that $\sigma_2$ does not depend on the target nucleus 
 then $\sigma_2 (A,x,Q^2) \approx $ $\sigma^D(x,Q^2)$  and $\frac{A}{2} a_2(A/d)$ can be connected to the number 
 of correlated pairs in $A$ \cite{Vanhalst:2012ur}. It is clear that the SRC contain no explicit quark-gluon effects and arise
solely from nucleonic dynamics.


\begin{figure*}[htb]
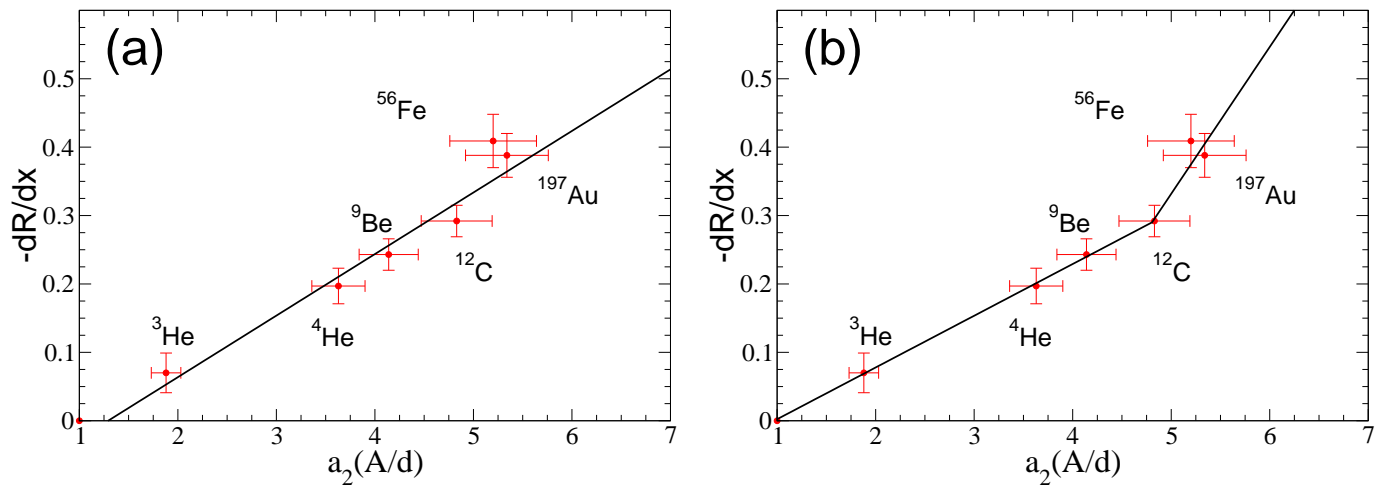

\resizebox{\textwidth}{!}{
  \includegraphics[clip]{EMCvsA2.eps}  \hskip 1.0cm \includegraphics[clip]{EMCvsA2-2.eps}}
\caption{The EMC slopes extracted from Refs. \cite{Weinstein:2010rt} versus the $a_2(A/d)$ parameters extracted from \cite{Frankfurt:1993sp,Egiyan:2005hs,Fomin:2011ng} as described in Table I of Ref.  \cite{Hen:2012fm} and the result of the fitting with (a)a linear function (b) a piecewise linear function consisting of two parts with different slopes.}
\label{dRdxa2}
\end{figure*}
%

Let us show the mentioned dependence of the EMC effect by plotting its slope versus the SRC scale factor $a_2(A/d)$.  In Fig. \ref{dRdxa2}a
we plot the overall fit using the data of columns 2, 4, 7  and 8 of table I in Ref. \cite{Hen:2012fm}, which
correspond to data  from Refs. \cite{Frankfurt:1993sp}, \cite{Egiyan:2005hs} and \cite{Fomin:2011ng} for $a_2(A/d)$, and Ref. \cite{Weinstein:2010rt} for $dR/dx$.  
We have assigned one value for each nucleus which reflects the weighted average of different independent measurements and added a $5\% $ uncertainty to $a_2(A/d)$ due to the thereoretical corrections needed to extract the data. We obtain a reasonably good linear fit, with a slope of $0.090 \pm 0.012$ and a $\tilde{\chi}^2/dof =  2.31/4$,\footnote{
Note that in this analysis we include the errors in the $x$ and $y$ coordinates in the calculation of  $\tilde\chi^2 $. Let $x (i)$ and $y(i)$ be the data points and $\Delta x(i)$ and $\Delta y(i)$ the corresponding errors, let $y= a + b x $ be the fitted line, then $\tilde{\chi}^2 = \sum_i (y(i) -a -b x(i))^2/(\Delta y (i)^2 +( b \Delta x(i))^2)$.  Therefore, a good fit arises when $\tilde{\chi}^2/dof \leq 1$ \cite{numericalrecipes}. Note, that if  we would consider that the $x$ coordinate has no error, as is done in fits to experimental data like in Fig. 3, we would get  in this case $\chi^2/dof = 1.21$.}  but discover that the high $A$ nuclei lie all above, 
and most of low $A$ nuclei  below, the line.

\begin{figure}[h]
\resizebox{8.5 cm}{!}{%
  \includegraphics[clip]{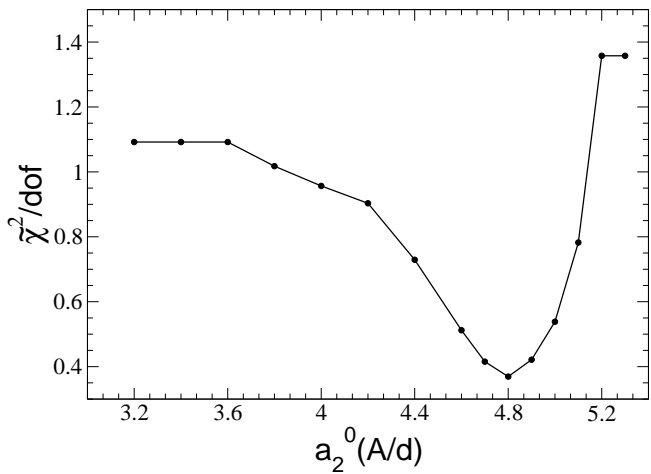}
}
\caption{$\tilde{\chi}^2$ divided by the number of degrees of freedom in the fit of the EMC slopes {\it vs.} $a_2(A/d)$ with a piecewise linear function as a function of parameter $a_2^0(A/d)$ that divides the region of light nuclei from heavy ones in Fig.\ref{dRdxa2}b.}
\label{fig3e}
\end{figure}

In the light of the latter observation we proceed to fit two lines one for heavy and the other for light nuclei. The position of the dividing point is obtained from minimization of $\tilde{\chi}^2$  and is found to correspond to $^{12}$C as shown in Fig. \ref{dRdxa2}b. In Fig. \ref{fig3e} we show the dependence of the $\tilde{\chi}^2$ on the position of the dividing point, it is clearly seen that the region of $^{12}$C corresponds to the minimum of the $\tilde{\chi}^2$.
The fit is now, $\tilde{\chi}^2/dof =0.152/3$,\footnote{With no errors for $a_2,  \chi^2/dof = 1.108/3$.} and  the slope of the straight line is steeper for heavy nuclei ($ 0.22 \pm 0.07$) than 
for light ones ($0.076 \pm 0.016$).  Therefore, we conclude that whatever physics governs these two observables has a drastic change around $^{12}$C. 
Regardless of this result, the fact that the EMC effect might be explained 
in terms of purely hadronic physics, is a matter of thought.

%

\begin{figure}[t]
\resizebox{8.5 cm}{!}{%
  \includegraphics[clip]{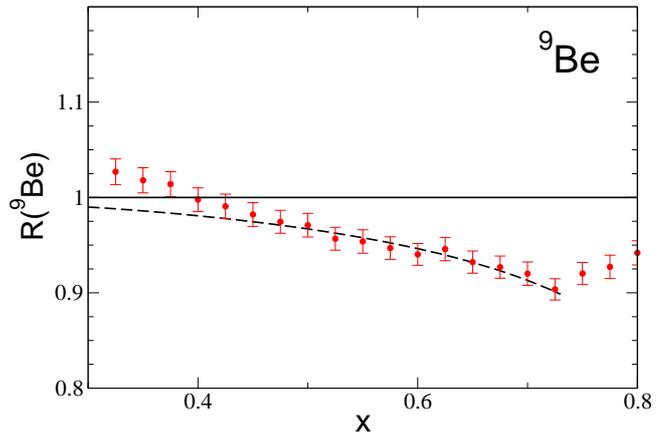}
}
\caption{Ratio of the $^9$Be  to the deuteron structure functions versus $x$, the proton scaling variable. Experimental data is of Ref.\cite{Seely:2009gt}. 
The dashed line corresponds to $\eta =1.0125$ in the formalism of Ref. \cite{GarciaCanal:1984eh}.}
\label{emc9Be}
\end{figure}
%
%
%

\begin{figure*}[!t]
\resizebox{\textwidth}{!}{
  \includegraphics[clip]{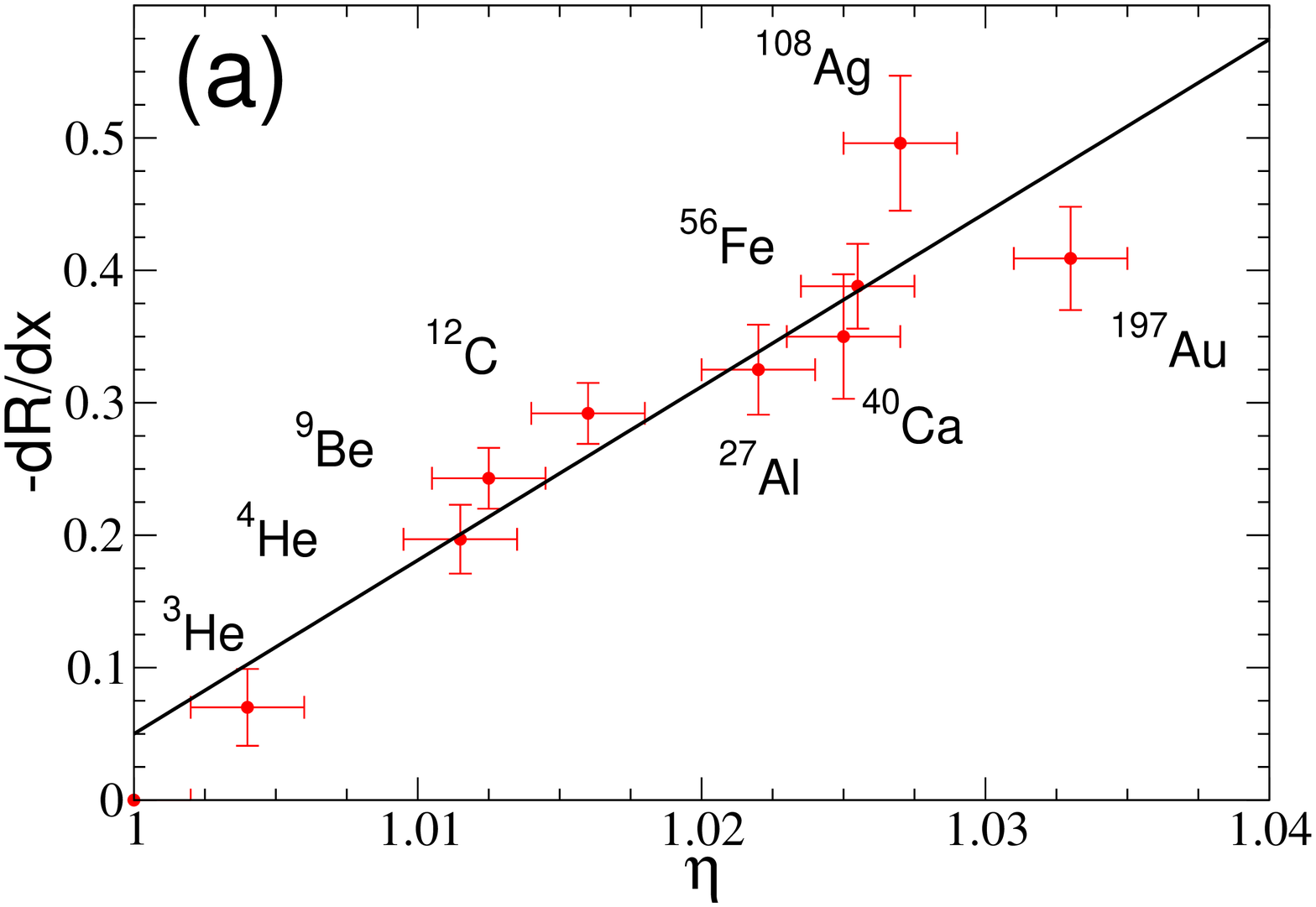}  \hskip 1.0cm \includegraphics[clip]{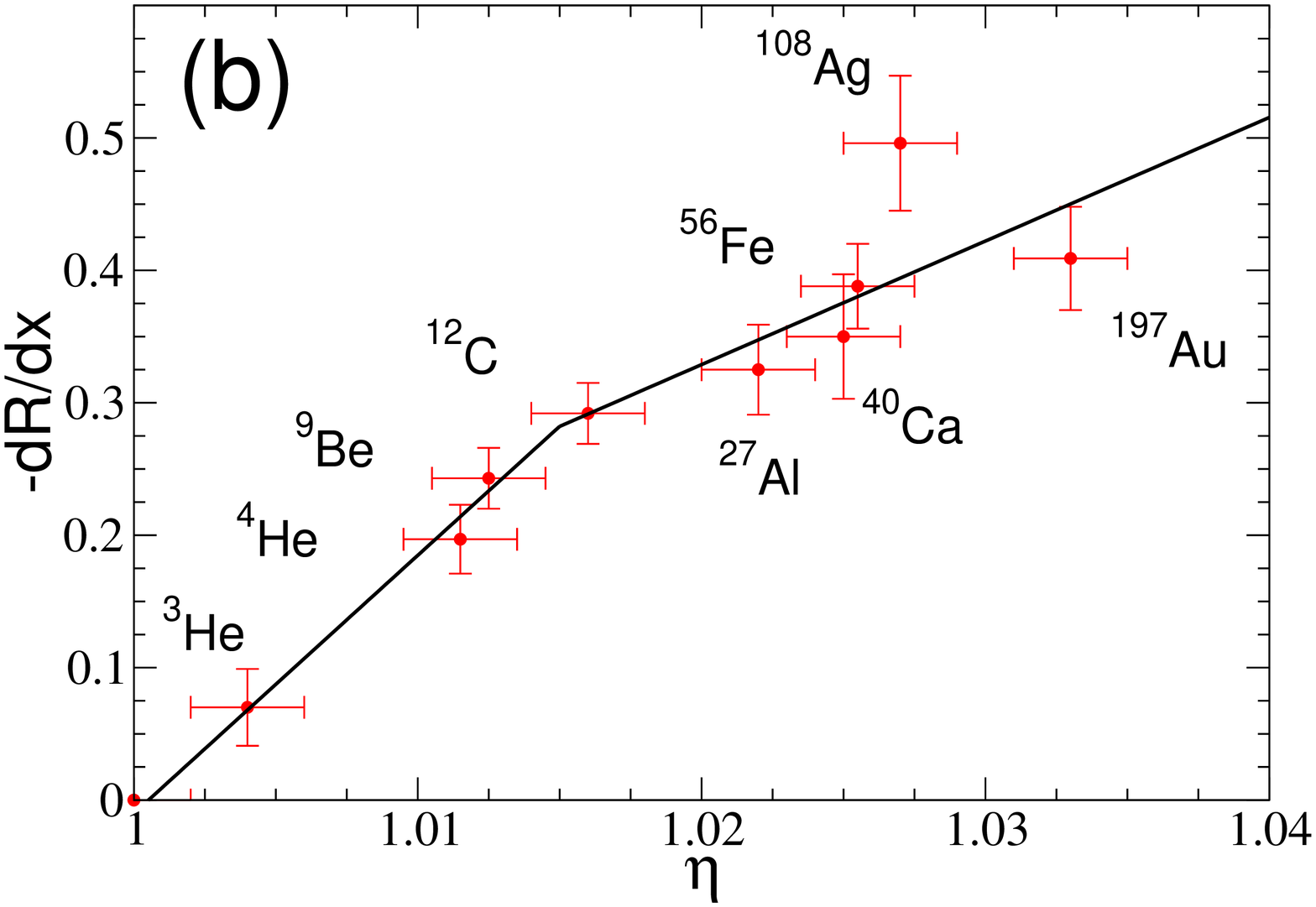}
}
\caption{The EMC slopes extracted from Refs. \cite{Weinstein:2010rt} versus the $x$-rescaling parameter $\eta$ and the result of the fitting with (a)a linear function b) a piecewise linear function consisting of two parts with different slopes.}
\label{dRdxeta}
\end{figure*}
%

%
\section{The Model}
\label{sec:1}

 These developments moved us to reanalyze the data in  a formalism with only one parameter per nucleus $\eta$ \cite{GarciaCanal:1984eh},  related to the binding energy of nucleons in nuclei \cite{Akulinichev:1985ij,GarciaCanal:1986xe}.  In this approach the EMC effect was described by suggesting that the true scaling variable for deep-inelastic scattering off nuclei should be taken to be
\begin{equation}
x^*=\eta x,
\end{equation}
with $0 < x < A$ and $\eta$ is the free parameter. 
Consequently, the ratios of the nuclear to deuteron structure functions can be written as follows

\begin{equation}
R(A)=F_2^{(A)}(x^*,Q^2)/F_2^{(D)}(x,Q^2),
\end{equation}
where $F_2^{(A)}(\tilde{x}^*,Q^2)$ is the nuclear structure function calculated using a rescaled variable $x^*$ and the free nucleon structure function, and $F_2^{(D)}(x,Q^2)$ is the deuteron structure function where the effects of rescaling are small. For the free nucleon structure functions we used the parametrization of Ref.\cite{Martin:2009iq}.

The main issue in that work was to obtain quark distributions shifted towards lower $x$ values  as compared to those corresponding to free nucleons, a mechanism which was called $x$-rescaling. 
A direct connection  between this $x$-rescaling approach and the $Q^2$-rescaling \cite{Close:1985ji}, based on the renormalization group 
evolution related to the perturbative quark-gluon structure  of the nucleons, was already shown in Ref.\cite{GarciaCanal:1986xe}. 

\begin{table}
\caption{For the measured nuclei the value of the $x$-rescaling parameter $\eta $ obtained by fitting the data of Ref. \cite{Arnold:1983mw,Seely:2009gt} 
and use the pdfs of Ref. \cite{Martin:2009iq} with our formalism. The value of the $x$-scaling parameter for $^2$H is assumed to be $\eta\approx 1$.}
\label{etaR}       
\begin{tabular}{|c|c|c|c|c|c|c|}
\hline\noalign{\smallskip}
A &   $^3$He &   $^4$He  & $^9$Be  & $^{12}$C  & $^{56}$Fe & $^{197}$Au\\
\noalign{\smallskip}\hline\noalign{\smallskip}
$\eta$& $1.0040 $ & $1.0115$ & $1.0125 $& $ 1.0160 $  & $ 1.0255$  & $1.0330$ \\

\noalign{\smallskip}\hline
\end{tabular}
\end{table}

\section{Analysis of data}
\label{sec:2}

The above formalism leads to fits of very good quality for the EMC effect in the region $0.3\leq x \leq 0.7$ based on the $x$-rescaling model. We show in Fig. \ref{emc9Be} an example of a standard fit
for  $^9$Be corresponding  to a value $\eta =1.0125$ restricted to the $x$-region of present interest. 
Similar quality was obtained for all the nuclei considered.
In Table \ref{etaR} we summarize the values of $\eta$ obtained.

In order to compare our results with the SRC proposal \cite{Weinstein:2010rt}, we take the values of the measured
EMC slopes they quote in their Table I, corresponding also to the range
$0.3\leq x \leq 0.7$. In addition to the nuclei presented in Fig.\ref{dRdxa2} and Table \ref{etaR} we use SLAC experimental data for the EMC ratios $R$ for $^{27}$Al, $^{40}$Ca and $^{108}$Ag \cite{Aubert:1983xm} and the corresponding $dR/dx$. We include these data to improve the statistics in the range of heavy nuclei and, as we discuss later, this allows us to present some predictions of the corresponding $a_2(A/d)$ values.

With these experimental data and the $\eta$
values obtained in the fit, we produce Fig. \ref{dRdxeta}a that also shows
a linear dependence of the EMC slopes with the effective mass
parameter $\eta$. The value of the slope is $13.11 \pm1.7$ and the quality of the fit is  characterized by $\tilde{\chi}^2/dof = 7.65/7$.  A similar fit in
terms of binding energy has been recently performed in Ref. \cite{Benhar:2012nj}.

We note by looking to Fig. \ref{dRdxeta}a  that, even if the linear fit with one straight line is quite reasonable,  a fit with a 
piecewise linear function consisting of two parts with different slopes, one for the light nuclei and one for 
the heavy as shown in Fig. \ref{dRdxeta}b, might be more precise. Our statistical analysis confirms our suspicion that the two line fit is at least as good as the one with one straight line.
In fact, we obtain $\tilde{\chi}^2/dof = 5.22/5$ and the corresponding 
slopes are  $19.5\pm 3.7$ for the light nuclei and $9.3\pm 2.5$ for the heavy ones. One sees that the slope for the heavy nuclei is two times smaller
than the slope for the light nuclei. Moreover, we realize that the transition from the region of light nuclei to that of the heavy nuclei takes place at $\eta=\eta^0=1.015$ which corresponds to carbon. This value of $\eta^0$ results from the minimization of $\tilde{\chi}^2$ in the fitting with a piecewise function.

\begin{figure}[h]
\resizebox{8.5 cm}{!}{%
  \includegraphics[clip]{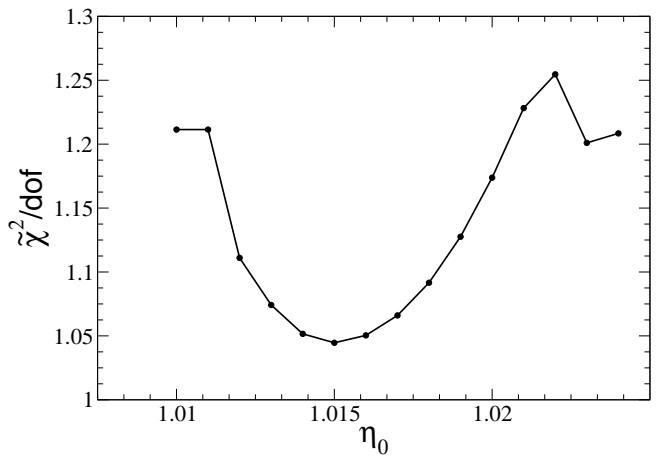}
}
\caption{$\tilde{\chi}^2$ divided by the number of degrees of freedom in the fit of the EMC slopes {\it vs.} $\eta$ with a piecewise linear function as a function of parameter $\eta^0$ that divides the region of light nuclei from heavy ones in Fig.\ref{dRdxeta}b and Fig.\ref{etaa2AL}.}
\label{chi2eta0}
\end{figure}

 In Fig. \ref{chi2eta0} we show  the calculated $\tilde{\chi}^2$ values (divided by the number of degrees of freedom)  as a function of
the parameter $\eta^0$ which is the value of the parameter $\eta$  that divides the range of $\eta$ into light and heavy nuclei. This value of $\tilde{\chi}^2$, comparable to the previous one, establishes that this fit might be physically motivated. The lack of data for heavy nuclei impedes a better discrimination.

Certainly, given the relation between all the various treatments  mentioned before, similar plots of $dR/dx$ as a function of 
confinement scales $R^*/R$ \cite{GarciaCanal:1986xe,Close:1985ji,Close:1984zn} can be produced.

We observe again that the physics one can associate with the heavy nuclei might be different from 
that related to light nuclei. The average nucleon interaction energy ($\sim 1/\eta$) grows faster with $A$ for the heavy nuclei than it does for the light nuclei \cite{GarciaCanal:1986xe,Moniz:1971mt}.

Thus we arrive to an impasse, we find linear correlations between $dR/dx$  both in the hadronic scheme  with $a_2(A/d)$,
and in the quark-gluon scheme with $\eta$. Both schemes are able to explain the EMC phenomenon ($0.3 < x <0.7$) and therefore 
we can associate it either to the effect of the  local nuclear environment as characterized by the
high momentum components of the nuclear wave function \cite{Seely:2009gt}, 
as well as, to $x$-rescaling through arguments based on perturbative QCD, but, and 
this is important, through the non perturbative parameter $\eta$ related to the the binding energy.

\begin{table}
\caption{ The values of the SRC scale factor $a_2(A/d)$ for $^{27}$Al, $^{40}$Ca and $^{108}$Ag using two different methods: (a) -- the first line -- using the parametrization of $a_2(A/d)$ {\it vs.} $\eta$ from Fig.\ref{etaa2AL} and (b) -- second line -- using the parametrization of $dR/dx$ {\it vs.} $a_2(A/d)$ from Fig.\ref{dRdxa2}b}
\label{a2predict}       
\begin{tabular}{|c|c|c|c|}
\hline\noalign{\smallskip}
A &   $^{27}$Al &   $^{40}$Ca  & $^{108}$Ag\\
\noalign{\smallskip}\hline\noalign{\smallskip}
$a_2(A/d)$ (Fig.\ref{etaa2AL}) & $5.05 \pm 0.07$ & $5.15\pm0.07$ & $5.22\pm0.07$  \\
$a_2(A/d)$ (Fig.\ref{dRdxa2} ) & $4.96\pm 0.15$ & $5.06\pm0.21$ & $5.68\pm 0.23$ \\
\noalign{\smallskip}\hline
\end{tabular}
\end{table}

We can establish a linear relation  between the parameter $\eta$ and the SRC scale factor $a_2(A/d)$, also found in ref. \cite{Vanhalst:2012zt} . However, in Fig. \ref{etaa2AL} we show that this relation is better fitted again by a two linear fit. This implies that both treatments are
related in a mathematical form which distinguishes between light and heavy nuclei.
This plot allows us to predict the values of the scale factor $a_2(A/d)$ for $^{27}$Al, $^{40}$Ca and $^{108}$Ag.
We present these values in the Table \ref{a2predict}, together with the $a_2(A/d)$ predictions
from the EMC slopes of the $dR/dx$  {\it vs.}  $a_2(A/d)$ parametrization in the region of heavy nuclei from the Fig.\ref{dRdxa2}b. 
The errors were estimated using the error of parameter $\eta$ for the former and using the known experimental error for EMC slopes  for the latter. It is seen that the values of $a_2(A/d)$ for $^{27}$Al and $^{40}$Ca agree within the error.

%

\section{Conclusions}
\label{sec:3}

Before we proceed to comment on these results we would like to emphasize that the main outcome of our investigation has been to make explicit that the interesting linear correlation between the EMC effect and the  SRC  can be extended to a linear correlation between the EMC effect and the $x$-rescaling parameter $\eta$ which leads to a linear correlation between $\eta$ and SRC, for us a signature of QCD-hadron duality. We  have also made the  observation that 
 these linear relations might be too naive and that some high-low $A$ average  dynamics might separate the EMC effect in two regimes.  More data should clarify this issue and help understand QCD-hadron duality.

However, none of the shown treatments is purely QCD or purely hadronic. Both contain in their parametrizations the other
component. In the case of the $x$-rescaling the hadronic behavior enters through  the $\eta$ parameter,
 an in medium hadronic property.
In the case of the SRC motivated relation one should recall Weinberg's theorem \cite{Weinberg:1978kz} which implies that QCD in some
kinematical regime can be understood in terms of purely hadronic degrees of freedom. The quark-gluon behavior arises through the fitting of the dynamical constants.
These treatments are both highly non additive. 

\begin{figure}[t]
\resizebox{8.5 cm}{!}{%
  \includegraphics[clip]{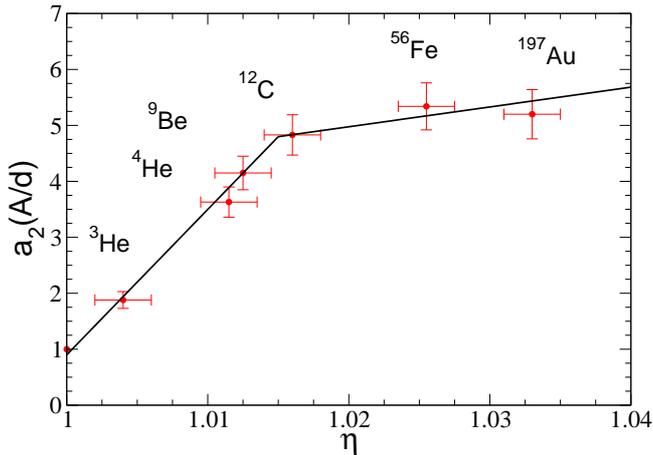}}
\caption{ The short range correlation scale factor $a_2(A/d)$ versus the nucleon $x$-rescaling parameter $\eta$. 
We see a low $A$ and a high $A$ linear relation.}
\label{etaa2AL}
\end{figure}

Fig. \ref{etaa2AL} shows a low variation of $a_2(A/d)$ with $\eta$ for large $A$, which
suggests a low sensitivity of $\eta$ to the detailed nuclear dynamics for heavy nuclei.  On the other hand, the rapid variation at 
low $A$, suggests that a QCD type description might be adequate in this regime. 
This fit has been made assuming the $\eta_0 =1.015$,  as obtained in Fig. \ref{chi2eta0}, and we obtain $\tilde{\chi^2}/dof = 0.524/3$ for it.

 It has been argued  that one should use $x_N = Q^2/2P_A q$, where $P_A$ is the nucleus four momentum, rather than  $x$ to represent the data in order  to work in the nucleus reference frame \cite{Frankfurt:2012qs}. This is precisely what Fig. \ref{etaa2AL} does. It represent the data in terms of the nucleus effective $x^* = \eta x$ and therefore the $A$ dependence is dynamical and not kinematical.

The $A$ dependence of the EMC contribution has been obtained 
in a microscopic treatment of the EMC effect which separates all the various components contributing to the process, i.e.  nucleon structure, equivalent photon 
 and the hadronic components \cite{Frankfurt:2012qs}. In our treatment the various dependences are implicit. Note that their scaling variable is purely perturbative
 not like our $\eta$ parameter which is non perturbative.  A relation between both approaches would clarify some issues and provide the $x$ dependence. 
 We  may speculate at this point based on preliminary calculations that the low $A$ and low $x$ ($0.3<x<0.5$) regime is best suited to see  the  almost perturbative $x$-rescaling behavior.

The  linear correlation between the different parameters shown in here  allows one to conclude that in hadronic language, as it was suggested in \cite{Seely:2009gt}, the
nuclear dependence of the quark distributions is directly related to the local nuclear environment of the acted nucleon, and  in terms of quark-gluon language the nucleons in the medium are bound and therefore their x-rescaling parameter $\eta$ is larger than in the vacuum. Both statements represent a dual view  of the EMC effect.

%
%
%

\begin{acknowledgement}
 We  acknowledge useful conversations with H. Fanchiotti. We thank O. Hen for a clarifying discussion regarding the linear fits.  We are thankful to J. Ryckebusch for sending their results prior to publication and for useful comments. C.A.G.C. acknowledges illuminating exchanges with  R. Sassot. C.A.G.C.  and T.T. have been partially supported  ANPCyT Argentina. V.V. has been funded by the Ministerio de Econom\'ia y Competitividad and  EU FEDER under contract FPA2010-21750-C02-01, by Consolider Ingenio 2010
CPAN (CSD2007-00042), by Generalitat Valenciana: Prometeo/2009/129, by the
European Integrated Infrastructure Initiative HadronPhysics3--(Grant number
283286).
\end{acknowledgement}


\begin{thebibliography}{60}
\bibitem{Aubert:1983xm} 
  J.~J.~Aubert {\it et al.}  [European Muon Collaboration],
  Phys.\ Lett.\ B {\bf 123}, (1983) 275.

\bibitem{Arnold:1983mw} 
  R.~G.~Arnold, P.~E.~Bosted, C.~C.~Chang, J.~Gomez, A.~T.~Katramatou, G.~Petratos, A.~A.~Rahbar and S.~Rock {\it et al.},
  Phys.\ Rev.\ Lett.\  {\bf 52}, (1984) 727.

\bibitem{Seely:2009gt}
  J.~Seely, A.~Daniel, D.~Gaskell, J.~Arrington, N.~Fomin, P.~Solvignon, R.~Asaturyan and F.~Benmokhtar {\it et al.},
  Phys.\ Rev.\ Lett.\  {\bf 103}, (2009) 202301.
  [arXiv:0904.4448 [nucl-ex]].

\bibitem{Higinbotham:2010tb}
  D.~W.~Higinbotham,
  AIP Conf.\ Proc.\  {\bf 1374}, (2011) 85.
  [arXiv:1010.4433 [nucl-ex]].

\bibitem{Weinstein:2010rt}
  L.~B.~Weinstein, E.~Piasetzky, D.~W.~Higinbotham, J.~Gomez, O.~Hen and R.~Shneor,
  Phys.\ Rev.\ Lett.\  {\bf 106}, (2011) 052301.
  [arXiv:1009.5666 [hep-ph]].

\bibitem{Piasetzky:2011zz}
  E.~Piasetzky, L.~B.~Weinstein, D.~W.~Higinbotham, J.~Gomez, O.~Hen and R.~Shneor,
  Nucl.\ Phys.\ A {\bf 855}, (2011) 245.
  
\bibitem{Arrington:2012ax}
  J.~Arrington, A.~Daniel, D.~Day, N.~Fomin, D.~Gaskell and P.~Solvignon,
  Phys.\ Rev.\ C {\bf 86} (2012) 065204
  [arXiv:1206.6343 [nucl-ex]].
  
\bibitem{Hen:2012fm} 
  O.~Hen, E.~Piasetzky and L.~B.~Weinstein,
  Phys.\ Rev.\ C {\bf 85}, (2012) 047301.
  [arXiv:1202.3452 [nucl-ex]].
  
  
  
\bibitem{Frankfurt:1993sp} 
  L.~L.~Frankfurt, M.~I.~Strikman, D.~B.~Day and M.~Sargsyan,
  Phys.\ Rev.\ C {\bf 48}, 2451 (1993).

\bibitem{Frankfurt:2009vv}
  L.~Frankfurt, M.~Sargsyan and M.~Strikman,
  AIP Conf.\ Proc.\  {\bf 1056}, (2008) 322.
  [arXiv:0901.2340 [nucl-th]].


\bibitem{Vanhalst:2012ur}
  M.~Vanhalst, J.~Ryckebusch and W.~Cosyn,
  Phys.\ Rev.\ C {\bf 86} (2012) 044619
  [arXiv:1206.5151 [nucl-th]].
  
\bibitem{Egiyan:2005hs} 
  K.~S.~Egiyan {\it et al.}  [CLAS Collaboration],
  Phys.\ Rev.\ Lett.\  {\bf 96}, (2006) 082501.
  [nucl-ex/0508026].
  
  
\bibitem{Fomin:2011ng} 
  N.~Fomin, J.~Arrington, R.~Asaturyan, F.~Benmokhtar, W.~Boeglin, P.~Bosted, A.~Bruell and M.~H.~S.~Bukhari {\it et al.},
  Phys.\ Rev.\ Lett.\  {\bf 108}, (2012) 092502.
  [arXiv:1107.3583 [nucl-ex]].
  
\bibitem{numericalrecipes}
W.~ H.~ Press, S.~ A.~ Teukolsky, W.~ T.~ Vetterling, B.~ P.~ Flannery,
{\it Numerical  Recipes: The Art of Scientific Computing}, Third Edition (2007), Cambridge University Press.
  
 
  
 
  
\bibitem{GarciaCanal:1984eh} 
  C.~A.~Garcia Canal, E.~M.~Santangelo and H.~Vucetich,
  Phys.\ Rev.\ Lett.\  {\bf 53}, (1984) 1430.

\bibitem{Akulinichev:1985ij} 
  S.~V.~Akulinichev, S.~A.~Kulagin and G.~M.~Vagradov,
  Phys.\ Lett.\ B {\bf 158}, (1985) 485.

\bibitem{GarciaCanal:1986xe} 
  C.~A.~Garcia Canal, E.~M.~Santangelo and H.~Vucetich,
  Phys.\ Rev.\ D {\bf 35}, (1987) 382.
  
\bibitem{Martin:2009iq} 
  A.~D.~Martin, W.~J.~Stirling, R.~S.~Thorne and G.~Watt,
  Eur.\ Phys.\ J.\ C {\bf 63}, (2009) 189.
  [arXiv:0901.0002 [hep-ph]].
  
\bibitem{Close:1985ji} 
  F.~E.~Close, R.~G.~Roberts and G.~G.~Ross,
  Phys.\ Lett.\ B {\bf 168}, (1986) 400.


\bibitem{Benhar:2012nj} 
  O.~Benhar and I.~Sick,
  arXiv:1207.4595 [nucl-th].


\bibitem{Close:1984zn} 
  F.~E.~Close, R.~L.~Jaffe, R.~G.~Roberts and G.~G.~Ross,
  Phys.\ Rev.\ D {\bf 31}, (1985) 1004.

\bibitem{Moniz:1971mt}
  E.~J.~Moniz, I.~Sick, R.~R.~Whitney, J.~R.~Ficenec, R.~D.~Kephart and W.~P.~Trower,
  Phys.\ Rev.\ Lett.\  {\bf 26} (1971) 445.
  
\bibitem{Vanhalst:2012zt}
  M.~Vanhalst, J.~Ryckebusch and W.~Cosyn,
  arXiv:1210.6175 [nucl-th].
 

\bibitem{Weinberg:1978kz} 
  S.~Weinberg,
  Physica A {\bf 96}, (1979) 327.

  
\bibitem{Frankfurt:2012qs}
  L.~Frankfurt and M.~Strikman,
  Int.\ J.\ Mod.\ Phys.\ E {\bf 21} (2012) 1230002
  [arXiv:1203.5278 [hep-ph]].
  




\end{thebibliography}
\end{document}